# High  Temperature  Superconductivity

M. Brian Maple

Department of Physics and Institute for Pure and Applied Physical Sciences

University of California, San Diego, La Jolla, California 92093, USA

## Abstract

The current status of basic research on the high temperature cuprate superconductors and prospects for technological applications of these materials is discussed.  Recent developments concerning other novel superconductors are also briefly described.

**Contact  author:**

M. Brian Maple, Department of Physics–0319, University of California, San Diego, 9500 Gilman Drive, La Jolla, California 92093-0319, USA; FAX: (619) 534-1241; email: mbmaple@ucsd.edu

**Introduction**

The discovery of superconductivity at ~30 K in the La-Ba-Cu-O system by Bednorz and Müller in 1986 [1] ignited an explosion of interest in high temperature superconductivity. These initial developments rapidly evolved into an intense worldwide research effort that still persists after more than a decade, fueled by the fact that high temperature superconductivity constitutes an extremely important and challenging intellectual problem and has enormous potential for technological applications. During the past decade of research on this subject, significant progress has been made on both the fundamental science and technological applications fronts. For example, the symmetry of the superconducting order parameter and the identity the superconducting electron pairing mechanism appear to be on the threshold of being established, and prototype superconducting wires that have current carrying capacities in high magnetic fields that satisfy the requirements for applications are being developed. Prospects seem to be good for attaining a fundamental understanding of high temperature superconductors and realizing technological applications of these materials on a broad scale during the next decade.

The purpose of this paper is to provide a brief overview of the current status of the field of high temperature superconductivity. The emphasis is on experiment and recent developments on the high superconducting critical temperature ($T_c$) cuprates. Topics discussed include: (1) materials, (2) structure and charge carrier doping, (3) normal state properties, (4) symmetry of the superconducting order parameter, and (5) prospects for technological applications. At the end of the article, we describe recent progress involving other novel superconducting materials.

The immense scope of this subject dictated a very selective choice of the examples cited to illustrate the progress made in this field over the past decade. For comprehensive accounts of specific topics in high temperature superconductivity, the reader is referred to various review articles, such as those that appear in the series of volumes edited by Ginsberg [2].  Because of space limitations, it was also not possible to discuss the fascinating subject of vortex phases and dynamics which has flourished since the discovery of the cuprate superconductors [3].

**High $T_c$ superconducting cuprates**

The dramatic increases in $T_c$ that have been observed since 1986 are illustrated in Fig. 1 where the maximum value of $T_c$ is plotted vs. date.  Prior to 1986, the A15 compound $Nb_3Ge$ with $T_c \approx 23$ K held the record for the highest value of $T_c$ [4].  The maximum value of $T_c$ has increased steadily since 1986 to its present value of ~133 K for a compound in the Hg-Ba-Ca-Cu-O system [5,6].  When this compound ($HgBa_2Ca_2Cu_3O_8$) is subjected to a high pressure, the $T_c$ onset increases to ~164 K (more than half way to room temperature!) at pressures ~30 GPa [7,8].  While $HgBa_2Ca_2Cu_3O_8$ cannot be used in applications of superconductivity at such high pressures, this striking result suggests that values of $T_c$ in the neighborhood of 160 K, or even higher, are attainable in cuprates at atmospheric pressure.

Values of $T_c$ in excess of the boiling temperature of liquid nitrogen (77 K) immediately implicated high $T_c$ cuprates as promising candidates for technological applications of superconductivity.  Whereas liquid helium is currently employed to cool conventional superconducting materials such as Nb, NbTi, and $Nb_3Sn$ into the superconducting state (Nb is employed in SQUIDs and NbTi and $Nb_3Sn$ are used to make superconducting wires), the cuprate materials have the advantage that they can be cooled into the superconducting state using liquid nitrogen.  Cuprates such as the $LnBa_2Cu_3O_{7-\delta}$ (Ln = lanthanide) compounds ($T_c$ in the range 92-95 K) have enormous critical fields ~$10^2$ tesla [9,10] that are more than adequate for technological applications.  Epitaxially grown thin films of $YBa_2Cu_3O_{7-\delta}$ on single crystal $SrTiO_3$ substrates have critical current densities $J_c \approx 10^6 A/cm^2$ in zero field which decrease relatively slowly with magnetic field, making them suitable for technological applications [11].  Unfortunately, polycrystalline bulk materials have $J_c$'s that are disappointingly low, ~$10^3$ - $10^4$ A/cm, and are strongly depressed by a magnetic field [12]. The situation can be improved substantially by subjecting $YBa_2Cu_3O_{7-\delta}$ to a melt textured growth process which yields values of $J_c$ of $\sim 10^5$ A/cm at 77 K that are not too strongly depressed by an applied magnetic field [13]. Fortunately, techniques have recently been devised which yield values of $J_c$ in high fields for in-plane grain oriented thin films of $YBa_2Cu_3O_{7-\delta}$ on flexible substrates at 64 K (pumped liquid nitrogen temperatures) that exceed those of NbTi and $Nb_3Sn$ at liquid helium temperatures [14]. These promising developments are briefly described near the end of this article.

**The materials**

Approximately 100 different cuprate materials, many of which are superconducting, have been discovered since 1986. Several of the more important high $T_c$ cuprate superconductors are listed in Table 1, along with the maximum values of $T_c$ observed in each class of materials. Included in the table are examples of abbreviated designations (nicknames) for specific cuprate materials which we will use throughout this article (e.g., $YBa_2Cu_3O_{7-\delta}$ = YBCO, YBCO-123. Y-123). High quality polycrystalline, single crystal and thin film specimens of these materials have been prepared and investigated extensively to determine their fundamental normal and superconducting state properties. Presently, two of the leading candidates for technological applications of superconductivity are the $LnBa_2Cu_3O_{7-\delta}$ and $Bi_2Sr_2Ca_2Cu_3O_{10}$ materials.

**Structure and charge carrier doping**

The high $T_c$ cuprate superconductors have layered perovskite-like crystal structures which consist of conducting $CuO_2$ planes separated by layers comprised of other elements A and oxygen, $A_mO_n$, and, in some cases, layers of Ln ions [15,16]. The mobile charge carriers, which can be electrons but are usually holes, are believed to reside primarily within the $CuO_2$ planes. The $A_mO_n$ layers apparently function as charge reservoirs and control the doping of the $CuO_2$ planes with charge carriers. In several of the compounds containing Ln layers, the Ln ions with partially-filled 4f electron shells and magnetic moments have been found to order antiferromagnetically at low temperature [9,12].

Many of the cuprates can be doped with charge carriers and rendered superconducting by substitution of appropriate elements into an antiferromagnetic insulating parent compound. For example, substitution of divalent Sr for trivalent La in the antiferromagnetic insulator $La_2CuO_4$ dopes the $CuO_2$ planes with mobile holes and produces superconductivity in $La_{2-x}Sr_xCuO_4$ with a maximum $T_c$ of ~40 K at $x \approx 0.17$ [17]. Similarly, substitution of tetravalent Ce for trivalent Nd in the antiferromagnetic insulating compound $Nd_2CuO_4$ apparently dopes the $CuO_2$ planes with electrons, resulting in superconductivity in $Nd_{2-x}Ce_xCuO_{4-y}$ with a maximum $T_c$ of ~25 K at $x \approx 0.15$ for $y \approx 0.02$ [18,19]. The temperature T vs x phase diagrams for the $La_{2-x}Sr_xCuO_4$ and $Nd_{2-x}Ce_xCuO_{4-y}$ systems are shown in Fig. 2 [19], and the corresponding crystal structures of the $La_2CuO_4$ and $Nd_2CuO_4$ parent compounds are displayed in Fig. 3. The $La_{2-x}Sr_xCuO_4$ and $Nd_{2-x}Ce_xCuO_{4-y}$ systems have one $CuO_2$ plane per unit cell and are referred to as single $CuO_2$ layer compounds. Other superconducting cuprate systems have more than one $CuO_2$ plane per unit cell: $LnBa_2Cu_3O_{7-\delta}$ has two $CuO_2$ planes per unit cell (double $CuO_2$ layer compound), while $Bi_2Sr_2Ca_{n-1}Cu_nO_x$ has n $CuO_2$ layers per unit cell (n $CuO_2$ layer compound) and can be synthesized by conventional methods for n = 1, 2, 3.

Two features in Fig. 2 would appear to be relevant to cuprate superconductivity: (1) the apparent electron-hole symmetry may provide a constraint on viable theories of high $T_c$ superconductivity in cuprates, and (2) the proximity of antiferromagnetism suggests that superconducting electron pairing in the cuprates may be mediated by antiferromagnetic spin fluctuations. An antiferromagnetic pairing mechanism is consistent with the occurrence of d-wave pairing with $d_{x^2-y^2}$ symmetry that is suggested by experiments on several hole-doped cuprates (discussed later in this article). A number of theoretical models (e.g., [20-22]) based on AFM spin fluctuations predict d-wave superconductivity with $d_{x^2-y^2}$ symmetry for the cuprates. Surprisingly, as discussed below, experiments on the electron-doped superconductor $Nd_{2-x}Ce_xCuO_{4-y}$ suggest s-wave pairing similar to that of conventional superconductors where the pairing is mediated by phonons.

It is interesting that superconductivity with values of $T_c$ in the neighborhood of 30 K have been found in two noncuprate materials: the cubic perovskite $Ba_{1-x}K_xBiO_3$ ($T_c \approx 30$ K) [23,24] and the fcc "buckeyball" compound $Rb_3C_{60}$ ($T_c \approx 29$ K) [25,26]. Other features are consistent with non-phonon-mediated pairing in the hole-doped cuprates. The curve of $T_c$ vs carrier concentration can be approximated by an inverted parabola with the maximum value of $T_c$ occurring at an optimal dopant concentration $x_o$ [27]. (Note that the terminology "under-doped" refers to values of x smaller than the "optimally-doped" value $x_o$, whereas "over-doped" refers values of x larger than $x_o$.) The isotope effect on $T_c$ for optimally-doped material is essentially zero (i.e., $T_c \alpha M^{-\alpha}$ with $\alpha \approx 0$; M = ion mass) [28].

**Normal state properties**

It was realized at the outset that the normal state properties of the high $T_c$ cuprate superconductors are unusual and appear to violate the Landau Fermi liquid paradigm [29-32]. Some researchers share the view that in so far as the normal state properties reflect the electronic structure that underlies high $T_c$ superconductivity, it will be necessary to develop an understanding of the normal state before the superconducting state can be understood.

The anomalous normal state properties first identified in the high $T_c$ cuprate superconductors include the electrical resistivity and Hall effect. The electrical resistivity $\rho_{ab}(T)$ in the ab-plane of many of the hole-doped cuprate superconductors near optimal doping has a linear temperature dependence between $T_c$ and high temperatures ~1000 K, with an extrapolated residual resistivity $\rho_{ab}(0)$ that is very small; i.e., $\rho_{ab}(T) \approx \rho_{ab}(0) + cT$, with $\rho_{ab}(0) \approx 0$ and the value of c similar within different classes of cuprate materials [33]. The Hall coefficient $R_H$ is inversely proportional to T and the cotangent of the Hall angle $\theta_H = R_H/\rho$ varies as $T^2$; i.e., $\cot(\theta_H) \equiv \sigma_{xx}/\sigma_{xy} = AT^2 + B$ [34]. The linear T-dependence of $\rho(T)$ and the quadratic T-dependence of $\cot(\theta_H)$ have been attributed to longitudinal and transverse scattering rates $\tau_l^{-1}$ and $\tau_t^{-1}$ that vary as T and $T^2$, respectively [35]. In the RVB model, the constant and $T^2$ terms in $\tau_t^{-1}$ and, in turn, $\cot(\theta_H)$, are ascribed to scattering of spinons by magnetic impurities and other spinons, respectively. Examples of the linear T-dependence of $\rho(T)$, inverse T-dependence of $R_H$, and quadratic T-dependence of $\cot(\theta_H)$ near optimal doping ($x \approx 0$) can be found in Figs. 4(a), 5(a), and 5(b) in which $\rho_{ab}$, $R_H^{-1}$, and $\cot(\theta_H)$ vs T data are displayed for the $Y_{1-x}Pr_xBa_2Cu_3O_{7-\delta}$ system [36-38]. Experiments in which $Ca^{2+}$ ions are counterdoped with Pr for Y in $YBa_2Cu_3O_{7-\delta}$ indicate that the Pr ions localize holes at a rate of ~one hole per substituted Pr ion [37]. Thus, as x is increased, the $Y_{1-x}Pr_xBa_2Cu_3O_{7-\delta}$ system becomes more and more under-doped and $T_c$ decreases, vanishing near the metal-insulator transition that occurs at $x_{mi} \approx 0.55$. Displayed in Fig. 4(b) is the T-x phase diagram for the $Y_{1-x}Pr_xBa_2Cu_3O_{7-\delta}$ system which reveals the behavior of $T_c(x)$ as well as the Néel temperatures $T_N(x)$ for AFM ordering of Cu and Pr magnetic moments. It has been argued that the depression of $T_c$ with x is primarily due to the decrease in the number of mobile holes with increasing Pr concentration, although magnetic pair breaking by Pr may also be involved [37,38]. In contrast, both $\rho_{ab}(T)$ and $\rho_c(T)$ of the optimally-doped electron-doped cuprate $Sm_{1.83}Ce_{0.17}CuO_{4-y}$ vary as $T^2$, indicative of three dimensional Fermi liquid behavior [39].

The evolution of the normal ground state of the cuprates as a function of dopant concentration is particularly interesting. This is reflected in the temperature dependences of the ab-plane and c-axis electrical resistivities $\rho_{ab}(T)$ and $\rho_c(T)$ [40]. Both $\rho_{ab}(T)$ and $\rho_c(T)$ exhibit insulating behavior (i.e., $d\rho/dT < 0$) in the under-doped region, $\rho_{ab}(T)$ is metallic (i.e., $d\rho/dT > 0$) and $\rho_c(T)$ is insulating or metallic in the optimally-doped region, depending on the system, and $\rho_{ab}(T)$ and $\rho_c(T)$ are both metallic in the over-doped region. The linear T-dependence of $\rho_{ab}(T)$ and the insulating behavior of $\rho_c(T)$ suggest two dimensional non Fermi liquid behavior near the optimally-doped region, whereas the metallic $\rho(T) \propto T^n$ with n > 1 reflects a tendency towards three dimensional Fermi liquid behavior in the over-doped region. Recent measurements in 61-tesla pulsed magnetic fields to quench the superconductivity have been particularly useful in elucidating the evolution of $\rho_{ab}(T)$ and $\rho_c(T)$ with dopant concentration in the $La_{2-x}Sr_xCuO_4$ system [41]. Both $\rho_{ab}(T)$ and $\rho_c(T)$ were found to exhibit -ln T divergences in the under-doped region, indicative of a three dimensional non Fermi liquid [42]. As an example of the evolution of $\rho_{ab}(T)$ and $\rho_c(T)$ with doping, we again refer to the $Y_{1-x}Pr_xBa_2Cu_3O_{7-\delta}$ system. Shown in Fig. 6 are $\rho_{ab}(T)$ and $\rho_c(T)$ data for $Y_{1-x}Pr_xBa_2Cu_3O_{7-\delta}$ single crystals in the range of Pr concentrations $0 \leq x \leq 0.55$ [43]. The following features in the $\rho_{ab}(T)$ and $\rho_c(T)$ data in Fig. 6 are evident: a nonmonotonic evolution of $\rho_c(T)$ with x, the transformation of both $\rho_{ab}(T)$ and $\rho_c(T)$ from metallic to semiconducting with x, and the coexistence of metallic $\rho_{ab}(T)$ and semiconducting $\rho_c(T)$ for a certain range of doping. The nonmonotonic variation of $\rho_c(T)$ with x in Fig. 6(b) can be described with a phenomenological model [43] which assumes that the c-axis conductivity takes place via incoherent elastic tunneling between $CuO_2$ bilayers and CuO chain layers with a gap in the energy spectrum of the CuO chains (solid lines in Fig. 6(b)).

Perhaps the most remarkable aspect of the normal state is the pseudogap in the charge and spin excitation spectra of under-doped cuprates [44,45]. The pseudogap has been inferred from features in various transport, magnetic, and thermal measurements including $\rho_{ab}(T)$ [46,47,48], $R_H(T)$ [49], thermoelectric power S(T) [50], NMR Knight shift K(T) [51], NMR spin-lattice relaxation rate $1/T_1(T)$ [52,53,54], magnetic susceptibility $\chi(T)$ [49], neutron scattering [55], and specific heat C(T) [56], as well as spectroscopic measurements such as infrared absorption [57,58,59] and angle resolved photoemission spectroscopy ARPES [60,61]. An example of the features in $\rho_{ab}(T)$ that are associated with the pseudogap can be seen in the $\rho_{ab}(T)$ data displayed in Figs. 4(a) and 6(a) for the $Y_{1-x}Pr_xBa_2Cu_3O_{7-\delta}$ system. As the system becomes more under-doped with increasing x, $\rho_{ab}(T)$ deviates from linear behavior at higher temperature at a characteristic temperature T* which represents a crossover into the pseudogap state at T < T*. The transport, thermal, magnetic, and infrared studies of the pseudogap have been carried out on several cuprate materials, including LSCO, YBCO-123, YBCO-124, and BSCCO-2212, while the ARPES investigations of the pseudogap have mainly focussed on BSCCO, although ARPES measurements have also been made on oxygen-deficient YBCO.

The ARPES measurements reveal several striking aspects of the pseudogap. The magnitude of the pseudogap has the same k-dependence in the ab-plane as the magnitude of the superconducting energy gap, with maxima in the directions of $k_x$ and $k_y$ and minima at $45°$ to these directions. In fact, the symmetry is consistent with $d_{x^2-y^2}$ symmetry inferred from Josephson tunneling measurements on hole-doped cuprates discussed in the next section. Furthermore, measurements of the temperature dependence of the pseudogap at the angles where it is a maximum show that the superconducting gap grows continuously out of the pseudogap and that the value of the sum of both gaps at low temperatures is constant, independent of the temperature T* at which the pseudogap opens, or $T_c$. This is in marked contrast to the situation in conventional superconductors where the energy gap is proportional to $T_c$. These features are illustrated in Figs. 7(a) and (b) which show the k-dependence of the energy gap and the dependence of the maximum gap on temperature from the ARPES measurements of Ding et al. [60] on BSCCO-2212 samples wirh $T_c$'s of 87 K, 83 K, and 10 K. The pseudogap and the superconducting energy gap appear to be intimately related to one another, with the former the precursor of the latter. These results support the view that a unified theory of both the normal and superconducting states of the cuprates is imperative. The pseudogap and superconducting regions for the BSCCO-2212 derived from the ARPES measurements of Ding et al. [60] are shown in Fig. 8. Based upon investigations on LSCO, YBCO-123 and -124, BSCCO, and other systems, one can construct a generic T-x phase diagram which is shown schematically in Fig. 9. The phase diagram is very rich and contains insulating, antiferromagnetic, superconducting, pseudogap, two dimensional (2D) non Fermi liquid like, and three dimensional (3D) Fermi liquid like regions.

A number of models and notions have been proposed to explain the part of the phase diagram delineated by the curves of T* and $T_c$ vs x (e.g., [62–67]). Generally, these models involve the local pairing of electrons (or holes) at the temperature T* leading to a suppression of the low lying charge and spin excitations and the formation of the normal state pseudogap, followed by the onset of phase coherence at $T_c$ that results in superconductivity. Since the phenomenon of superconductivity involves coherent pairing, a bell shaped curve of $T_c$ vs x results as shown schematically in Fig. 10. For example, in resonating valence bond (RVB) models [30,62,63,64,67] which incorporate spin - charge separation into "spinons" with spin s = 1/2 and charge q = 0 and "holons" with s = 0 and q = +e, where e is the charge of the electron, the spinons become paired (spin pseudogap) at T* and coherent pairing of holons (Bose-Einstein condensation) occurs at $T_d$, resulting in superconductivity.

**Symmetry of the superconducting order parameter**

During the last several years, a great deal of effort has been expended to determine the symmetry of the superconducting order parameter of the high $T_c$ cuprate superconductors [68,69]. The pairing symmetry provides clues to the identity of the superconducting pairing mechanism which is essential for the development of the theory of high temperature superconductivity in the cuprates.

Shortly after the discovery of high $T_c$ superconductivity in the cuprates, it was established from flux quantization, Andreev reflection, Josephson effect, and NMR Knight-shift measurements, that the superconductivity involves electrons that are paired in singlet spin states [70]. Possible orbital pairing states include s-wave, extended s-wave, and d-wave states. In the s-wave state, the energy gap $\Delta(\mathbf{k})$ is <u>isotropic</u>; i.e., $\Delta(\mathbf{k})$ is constant over the Fermi surface. This leads to "activated" behavior of the physical properties in the superconducting state for T << $\Delta$; e.g., the specific heat $C_e(T)$, ultrasonic attenuation coefficient $\alpha_s(T)$, and NMR spin lattice relaxation rate $1/T_1(T)$ vary as exp $(-\Delta/T)$. For the extended s-wave state, the energy gap $\Delta(T)$ is <u>anisotropic</u>; i.e., $\Delta(\mathbf{k})$ exhibits a variation over the Fermi surface which has the same symmetry as the rotational symmetry of the crystal. Similarly, for the d-wave case, the energy gap $\Delta(\mathbf{k})$ is <u>anisotropic</u>, but with a symmetry that is <u>lower</u> than the symmetry of the crystal. The d-wave state that is consistent with most of the experiments discussed below has $d_{x^2-y^2}$ symmetry, which can be expressed as $\Delta(\mathbf{k}) = \Delta_o[\cos(k_x a) - \cos(k_y a)]$. For both the extended s-wave and d-wave cases, $\Delta(\mathbf{k})$ vanishes at lines on the Fermi surface, resulting in a density of states $N_s(E)$ that is linear in energy E at low values of E. This leads to "power law" $T^n$ (n = integer) behavior of the physical properties for $T << \Delta$; e.g., $C_e(T) \sim T^2$, the superconducting penetration depth $\lambda(T) \sim T$, and $1/T_1(T) \sim T^3$. The establishment of the symmetry of the superconducting order parameter requires the determination of both the <u>magnitude</u> and the <u>phase</u> of $\Delta(\mathbf{k})$. Shown in Fig. 11 is a schematic diagram of the variation of the energy gap over the Fermi surface and the density of states N(E) vs E for the "s", extended "s", and "d"$_{x^2-y^2}$ states [68].

A number of different types of measurements have been performed on the high $T_c$ cuprate superconductors which are sensitive to $|\Delta(\mathbf{k})|$. These include microwave penetration depth $\lambda(T)$ [71], microwave surface conductivity [72], NMR relaxation rate $1/T_1(T)$ [73], magnetic field dependence of the electronic specific heat $C_e(T)$ [74], thermal conductivity [75], ARPES [76], quasiparticle tunneling [77,78], and Raman scattering [79]. The results of these studies are generally consistent with a superconducting state with $d_{x^2-y^2}$ or extended s-wave symmetry for the hole-doped cuprates such as YBCO-123, 124, LSCCO, and BSCCO, and s-wave symmetry for the electron-doped superconductor NCCO [80,81,82]. However, the experiments discussed below indicate a dominant component of $d_{x^2-y^2}$ symmetry in the superconducting order parameter of hole-doped cuprates. This is a rather surprising result, considering the similarities in the structures of the hole-doped and electron-doped superconductors and the fact that they are both derived from chemical substitution of antiferromagnetic insulating parent compounds.

Several different types of measurements which are sensitive to the phase of $\Delta(\mathbf{k})$ have been performed. These measurements, all of which involve the Josephson effect, include SQUID interferometry [83,84,85], single junction modulation [86,87], tricrystal ring magnetometry [88,89,90], c-axis Josephson tunneling [91,92,93,94,95], and grain boundary tunneling [96]. The SQUID interferometry, single junction modulation, and tricrystal ring magnetometry measurements were performed on YBCO, while tricrystal magnetometry experiments have also been carried out on TBCCO. These experiments indicate that the superconducting order parameter in the YBCO and TBCCO hole-doped materials has $d_{x^2-y^2}$ symmetry. However, c-axis Josephson tunneling studies on junctions consisting of a conventional superconductor (Pb) and twinned or untwinned single crystals of YBCO indicate that the superconducting order parameter of YBCO has a significant s-wave component [91,92]. Shown in Fig. 12 are Josephson critical current vs magnetic field B data for an untwinned YBCO/Pb tunnel junction at 4.2 K for B||a and B||b [92]. From the geometry of the junction, independent measurements of the penetration depth of Pb, and the period of the $I_c(B)$ Fraunhofer patterns, the YBCO penetration depths $\lambda_a = 1307$ Å and $\lambda_b = 1942$ Å were obtained. The large anisotropy ratio $(\lambda_a/\lambda_b)^2 \leq 2$ suggests significant pair condensation on the Cu-O chains in YBCO.

Recently, a new class of c-axis Josephson tunneling experiments in which a conventional superconductor (Pb) was deposited across a single twin boundary of a YBCO single crystal were performed by Kouznetsov et al. [97]. The Josephson critical current $I_c$ was then measured as a function of magnitude and angle $\phi$ of magnetic field applied in the plane of the sample. For B aligned perpendicular to the twin boundary, a maximum in $I_c$ as a function of B was observed at B = 0, similar to the case of the untwinned YBCO single crystal data shown in Fig. 12, whereas for B parallel to the twin boundary, a minimum in $I_c$ was observed at B = 0. In all samples investigated, a clear experimental signature of an order parameter phase shift across the twin boundary was observed. The results provide strong evidence for mixed d- and s-wave pairing in YBCO and are consistent with predominant d-wave pairing with $d_{x^2-y^2}$ symmetry and a sign reversal of the s-wave component across the twin boundary.

An experiment that provides evidence for a multicomponent superconducting order parameter in YBCO was recently reported by Srikanth et al [98]. Microwave measurements on YBCO single crystals prepared in $BaZrO_3$ (BZO) crucibles [99] yielded new features in the temperature dependence of the microwave surface resistance and penetration depth in the temperature range $0.6 \leq T/T_c \leq 1$ which the authors suggest constitute evidence for a multicomponent superconducting order parameter. These features were not observed in microwave measurements on YBCO single crystals prepared with yttria stabilized zirconia (YSZ) crucibles, suggesting that YBCO single crystals prepared in BZO crucibles are of higher quality than those prepared in YSZ crucibles [99].

**Technological Applications of Superconductivity**

Technological applications of superconductivity can be divided into two major areas: superconducting electronics and superconducting wires and tapes. While the widespread use of high $T_c$ cuprate superconductors in technology has not yet been realized, steady and significant progress has been made towards this objective during the past decade. Recent developments indicate that high $T_c$ cuprate superconductors will begin to have a significant impact on technology during the next 5 to 10 years. The applications in superconducting electronics that are likely to be realized on this time scale have been summarized in a recent article by Rowell [100]. In the order in which they are anticipated, these applications include: SQUIDS, NMR coils, wireless communications subsystems, MRI coils and NMR microscopes, and digital instruments. In the area of superconducting wires and tapes, applications that appear to be feasible within this same time period include: power transmission lines, motors and generators, transformers, current limiters, magnetic energy storage, magnetic separation, research magnet systems, and current leads.

An example of recent progress in the area of superconducting wires and tapes is the development of flexible superconducting ribbons consisting of deposits of YBCO on textured substrates which have critical current densities $J_c \sim 10^6 A/cm^2$ in fields up to 8T at 64 K, a temperature that can be achieved by pumping on liquid nitrogen [14]. The performance of these prototype conductors in strong magnetic fields already surpasses that of NbTi and $Nb_3Sn$, which are currently used in commercial superconducting wires at liquid helium temperatures, in a comparable field range. The YBCO tapes are based on processes developed at two U. S. Department of Energy (DOE) National Laboratories, Los Alamos (LANL) and Oak Ridge (ORNL), under the DOE's National Superconductivity Program for Electric Power Applications. The essential step in both the LANL and ORNL processes is the preparation of a textured substrate, or "template," onto which a thick film of YBCO is deposited with well-aligned YBCO grains, that match the alignment of the underlying substrate. The alignment of the YBCO grains results in the high critical current density.

The LANL process, which is an extension of earlier work done at Fujikura Ltd. in Japan, uses a technique called "ion beam assisted deposition," or IBAD, to produce a preferentially oriented buffer layer on top of commercial nickel alloys, such as Hastelloy. The IBAD method uses four beams of charged particles to grow YSZ crystals with only one particular orientation on top of a ceramic oxide-buffered Hastelloy tape. Because of the excellent lattice match between YBCO and YSZ, the YBCO grains are "in-plane aligned" like the grains of the underlying YSZ layer.

The ORNL process involves fabricating long lengths of biaxially-textured metal (e.g., nickel) strips. Oxide buffer layers are then deposited on top of the metal substrate in order to transfer the alignment to the superconducting layer, while avoiding chemical degradation. The ORNL substrate technology is referred to as "RABiTS," or rolling-assisted, biaxially-textured substrates. In both the LANL and ORNL processes, pulsed laser deposition (PLD) is used to deposit the superconducting YBCO layer and some of the buffer layers.

Shown in Fig. 13 is the magnetic field dependence of the critical current density for the IBAD and RABiTS based YBCO coated samples recently produced at LANL and ORNL. These coated samples operated in the liquid nitrogen temperature range clearly outperform the metallic superconductors (NbTi, $Nb_3Sn$) at 4.2 K. Furthermore, even in the worst field direction (H ∥ c), and for temperatures below 65 K, the short sample YBCO coated conductors operated in a 8-tesla background field have at least an order of magnitude higher $J_c$ than pre-commercial BSCCO-2223 wire with no applied field.

The high $T_c$ YBCO coated tapes have been shown to be extremely flexible and to retain the high current carrying capacity, so that they appear to be suitable for wound magnet and coil applications. A significant challenge that remains is the development of efficient, continuous commercially viable processes for fabricating long lengths of these high current carrying in-plane aligned YBCO coated conductors with uniform properties.

**Other superconducting materials**

Although the high $T_c$ cuprates have been the focus of research on superconducting materials during the past decade, a number of other noteworthy novel superconducting materials have been discovered during this period. A few examples are briefly described below.

*Rare earth transition metal borocarbides*

Superconductivity was discovered in a series of compounds with the formula $RNi_2B_2C$ with a maximum $T_c$ of 16.5 K for R = Lu [101,102]. These materials have attracted a great deal of interest because they display both superconductivity and magnetic order and effects associated with the interplay of these two phenomena, similar to the $RRh_4B_4$ and $RMo_6X_8$ (X = S, Se) ternary superconductors that were studied extensively during the '70's and early '80's [103]. Investigations of superconducting and magnetic order and their interplay have been greatly facilitated by the availability of large single crystals of these materials [104]. Recently, values of $T_c$ that rival the $T_c = 23$ K value of the intermetallic compound $Nb_3Ge$, the high $T_c$ record holder for an intermetallic compound, have been found in mixed phase materials of the composition $YPd_5B_3C$ ($T_c = 23$ K) [105] and $ThPd_3B_3C$ ($T_c = 21$ K, $H_{c2}(0) \approx 17$ T) [106].

*$Sr_2RuO_4$*

The superconducting compound $Sr_2RuO_4$ has the same structure as the $La_{2-x}M_xCuO_4$ (M = Ba, Sr, Ca; Na) high $T_c$ cuprate superconductors [107]. While the $T_c$ of $Sr_2RuO_4$ is only ~1K, this compound is of considerable interest because it is the only layered perovskite superconductor without Cu. The anisotropy of the superconducting properties of $Sr_2RuO_4$ is very large ($\gamma = \xi_{ab}/\xi_c \approx 26$). Although this anisotropy is larger than that of $La_{2-x}M_xCuO_4$, the in-plane and c-axis resistivities of $Sr_2RuO_4$ vary as $T^2$ at low temperature, indicative of Fermi liquid behavior. It has been suggested that $Sr_2RuO_4$ may exhibit p-wave superconductivity [108].

*Alkali-metal doped $C_{60}$*

Exploration of the physical properties of materials based on the molecule $C_{60}$ revealed superconductivity with relatively high values of $T_c$ in metal-doped $C_{60}$ compounds [109]. For example, the fcc compounds $K_3C_{60}$ and $Rb_3C_{60}$ have $T_c$'s of 18 K and 29 K, respectively [25,26].

*$LiV_2O_4$*

The metallic transition metal oxide $LiV_2O_4$, which has the fcc normal-spinel structure, has been found to exhibit a crossover with decreasing temperature from localized moment to heavy Fermi liquid behavior [110], similar to that which has been observed in strongly correlated f-electron materials [111]. At 1 K, the electronic specific heat coefficient $\gamma \approx 0.42$ J/mol $K^2$, is exceptionally large for a transition metal compound. No superconducting or magnetic order was detected in this compound to temperatures as low as ~0.01 K. This behavior can be contrasted with that of the isostructural compound $LiTi_2O_4$ which displays nearly T-independent Pauli paramagnetism and superconductivity with $T_c$ = 13.7 K [112].

*Quantum Spin Ladder materials*

Quantum spin ladder materials have attracted much recent interest [113,114]. These materials consist of ladders made of AFM chains of S = 1/2 spins coupled by inter-chain AFM bonds. Examples of 2-leg ladder materials are $SrCu_2O_3$ and $LaCuO_{2.5}$; an example of a 3-leg ladder material is $Sr_2Cu_2O_5$. Superconductivity has apparently been discovered in the ladder material $Sr_{0.4}Ca_{13.6}Cu_{24}O_{41.84}$ under pressure with $T_c \approx$ 12 K at 3 GPa [115]. Part of the interest in quantum spin ladder materials stems from the fact that they are simple model systems for theories of superconductivity based on magnetic pairing mechanisms.

**Concluding remarks**

During the past decade, remarkable progress in the areas of basic research and technological applications has been made on the high $T_c$ cuprate superconductors. The availability of high quality polycrystalline and single crystal bulk and thin film materials has made it possible to make reliable measurements of the physical properties of these materials and to optimize superconducting properties (e.g., $J_c$) that are important for technological applications. These investigations have provided important information regarding the anomalous normal state properties, the symmetry of the superconducting order parameter, and vortex phases and dynamics in the cuprates. The next decade of research on the high $T_c$ cuprate superconductors as well as other novel superconducting materials promises to yield significant advances toward the development of a theory of high temperature superconductivity as well as the realization of technological applications of these materials on a broad scale. It is possible that significantly higher values of $T_c$ will be found in new cuprate compounds or other classes of materials. Nature may even have some more surprises in store for us, as it did in 1986!

**Acknowledgments**

Assistance in preparing the Plenary Lecture on which this paper is based was kindly provided by D. Basov, D. K. Christen, D. L. Cox, M. C. de Andrade, R. Chau, N. R. Dilley, R. C. Dynes, A. S. Katz, M. P. Maley, N. J. McLaughlin, and F. Weals.  This research was supported by the U. S. Department of Energy under Grant No. DE-FG03-86ER-45320.

## References

[1]     J. G. Bednorz and K. A. Müller, *Z. Phys. B*  64  (1986) 189.

[2]     *Physical Properties of High Temperature Superconductors I - V,*  D. M. Ginsberg, ed. (World Scientific, Singapore, 1989 - 1996).

[3]     See, for example, G. W. Crabtree and D. R. Nelson, *Physics Today*, April (1997) 38.

[4]     J. R. Gavaler, *Appl. Phys. Lett.* 23 (1973) 480.

[5]     S. N. Putilin, E. V. Antipov, O. Chmaissem, and M. Marezio, *Nature*  362 (1993) 226.

[6]     A. Schilling, M. Cantoni, J. D. Guo, and H. R. Ott, *Nature*  363 (1993) 56.

[7]     C. W. Chu, L. Gao, F. Chen, Z. J. Huang, R. L. Meng, and Y. Y. Xue, *Nature,* 365 (1993) 323.

[8]     M. Nuñez-Regueiro, J.-L. Tholence, E. V. Antipov, J.-J. Capponi, and M. Marezio, *Science*  262 (1993) 97.

[9]     M. B. Maple, Y. Dalichaouch, J. M. Ferreira, R. R. Hake, B. W. Lee, J. J. Neumeier, M. S. Torikachvili, K. N. Yang, H. Zhou, R. P. Guertin, and M. V. Kuric, *Physica B* 148 (1987) 155.

[10]    T. P. Orlando, K. A. Delin, S. Foner, E. J. McNiff, Jr., J. M. Tarascon, L. H. Greene, W. R. McKinnon, and G. W. Hull, *Phys. Rev. B* 36 (1987) 2394.

[11]    P. Chaudhari, R. H. Koch, R. B. Laibowitz, T. R. McGuire, and R. J. Gambino, *Phys. Rev. Lett.* 58 (1987) 2684.

[12]    For a review, see J. T. Markert, Y. Dalichaouch, and M. B. Maple, in *Physical Properties of High Temperature Superconductors I,* D. M. Ginsberg, ed. (World Scientific, Singapore, 1989), Ch. 6.

[13]    S. Jin, T. H. Tiefel, R. C. Sherwood, M. E. Davis, R. B. van Dover, G. W Kammlott, R. A. Fastnacht, and H. D. Keith, *Appl. Phys. Lett.* 52 (1988) 2074.

[14]    R. Hawsey and D. Peterson, *Superconductor Industry*, Fall, 1996; and references cited therein.

[15]    A. W. Sleight, *Physics Today*, June (1991) 24.

[16]    J. D. Jorgensen, *Physics Today*, June (1991) 34.

[17]    R. J. Cava, R. B. van Dover, B. Batlogg, and E. A. Rietman, *Phys. Rev. Lett.* 58 (1987) 408.

[18]    Y. Tokura, H. Takagi, and S. Uchida, *Nature*  337 (1989) 345.

[19]    M. B. Maple, *MRS Bulletin*  XV, No. 6 (1990) 60.

[20]    D. J. Scalapino, *Physics Reports*  250  (1995) 329; and references cited therein.

[21]    D. Pines, *Physica B* 199 & 200 (1994)  300; and references cited therein.

[22]    M. T. Béal-Monod and K. Maki, to be published.

[23]    L. F. Mattheis, E. M. Gyorgy, and D. W. Johnson, Jr., *Phys. Rev. B* 37 (1988) 3745.

[24]    R. J. Cava, B. Batlogg, J. J. Krajewski, R. Farrow, L. W. Rupp, Jr., A. E. White, K. Short, W. F. Peck, and T. Kometani, *Nature* 332 (1988) 814.

[25]    M. J. Rosseinsky, A. P. Ramirez, S. H. Glarum, D. W. Murphy, R. C. Haddon, A. F. Hebard, T. T. M. Palstra, A. R. Kortan, S. M. Zahurak, and A. V. Makhija, *Phys. Rev. Lett.* 66 (1991) 2830.

[26]    K. Holcer, O. Klein, S.-M. Huang, R. B. Kaner, K.-J. Fu, R. L. Whetten, and F. Diederich, *Science* 252 (1991) 1154.

[27]    S. Uchida, *Jpn. J. Appl. Phys.* 32 (1993) 3784 .

[28]    J. P. Franck, in *Physical Properties of High Temperature Superconductors IV,* D. M. Ginsberg, ed. (World Scientific, Singapore, 1994), p. 189.

[29]    See, for example, B. G. Levi, *Physics Today*, March (1990) 20; and references therein.

[30]    P. W. Anderson, *Science* 235 (1987) 1196.

[31]    R. B. Laughlin, *Science* 244 (1988) 525.

[32]    C. M. Varma, P. B. Littlewood, S. Schmitt-Rink, E. Abrahams, and A. E. Ruckenstein, *Phys. Rev. Lett.* 63 (1989) 1996.

[33]    See, for example, Y. Iye, in *Physical Properties of High Temperature Superconductors III*, D. M. Ginsberg, ed. (World Scientific, Singapore, 1992), Ch. 4.

[34]    T. R. Chien, Z. Z. Wang, and N. P. Ong, *Phys. Rev. Lett.* 67 (1991) 2088.

[35]    P. W. Anderson, *Phys. Rev. Lett.* 67 (1991) 2092.

[36]    M. B. Maple, C. C. Almasan, C. L. Seaman, S. H. Han, K. Yoshiara, M. Buchgeister, L. M. Paulius, B. W. Lee, D. A. Gajewski, R. F. Jardim, C. R. Fincher, Jr., G. B. Blanchet, and R. P. Guertin, *J. Superconductivity* 7 (1994) 97.

[37]    J. J. Neumeier, T. Bjornholm, M. B. Maple, and I. K. Schuller, *Phys. Rev. Lett.* 63 (1989) 2516.

[38]    J. J. Neumeier and M. B. Maple, *Physica C* 191 (1992) 158.

[39]    Y. Dalichaouch, C. L. Seaman, C. C. Almasan, M. C. de Andrade, H. Iwasaki, P. K. Tsai, and M. B. Maple, *Physica B* 171 (1991) 308.

[40]    See, for example, S. L. Cooper, and K. E. Gray, in *Physical Properties of High Temperature Superconductors IV*, D. M. Ginsberg, ed. (World Scientific, Singapore, 1994), Ch. 3.

[41]   G. S. Boebinger, Y. Ando, A. Passner, T. Kimura, M. Okuya, J. Shimoyama, K. Kishio, K. Tamasaku, N. Ichikawa, and S. Uchida, *Phys. Rev. Lett.* 77 (1996) 5417.

[42]   Y. Ando, G. S. Boebinger, A. Passner, T. Kimura, and K. Kishio, *Phys. Rev. Lett.* 75 (1995) 4662.

[43]   C. N. Jiang, A. R. Baldwin, G. A. Levin, T. Stein, C. C. Almasan, D. A. Gajewski, S. H. Han, and M. B. Maple, *Phys. Rev. B* 55 (1997) R3390.

[44]   See, for example, B. G. Levi, *Physics Today*, January (1996) 19; and references therein.

[45]   K. Levin, J. H. Kim,. J. P. Lu, and Q. Si, *Physica C* 175 (1991) 449.

[46]   B. Bucher, P. Steiner, J. Karpinski, E. Kaldis, and P. Wachter, *Phys. Rev. Lett.* 70 (1993) 2012.

[47]   B. Batlogg, H. Y. Hwang, H. Takagi, R. J. Cava, H. L. Rao, and J. Kwo, *Physica* 235-240C (1994) 130; and references therein.

[48]   T. Ito, K. Takenaka, and S. Uchida, *Phys. Rev. Lett.* 70 (1993) 3995.

[49]   H. Y. Hwang, B. Batlogg, H. Takagi, H. L. Kao, J. Kwo, R. J. Cava, J. J. Krajewski, and W. F. Peck, Jr., *Phys. Rev. Lett.* 72 (1994) 2636.

[50]   J. L. Tallon, J. R. Cooper, P. deSilva, G. V. M. Williams, and J. W. Loram, *Phys. Rev. Lett.* 75 (1995) 4114.

[51]   W. W. Warren, Jr., R. E. Walstedt, G. F. Brennert, R. J. Cava, R. Tycko, R. F. Bell, and G. Dabbagh, *Phys. Rev. Lett.* 62 (1989) 1193.

[52]   M. Takigawa, A. P. Reyes, P. C. Hammel, J. D. Thompson, R. H. Heffner, Z. Fisk, and K. C. Ott, *Phys. Rev. B* 43 (1991) 247.

[53]   H. Alloul, A. Mahajan, H. Casalta, and O. Klein, *Phys. Rev. Lett.* 70 (1993) 1171.

[54]   T. Imai, T. Shimizu, H. Yasuoka, Y. Ueda, and K. Kosuge, *J. Phys. Soc. Japan* 57 (1988) 2280.

[55]   J. Rossat-Mignod, L. P. Regnault, C. Vettier, P. Bourges, P. Burlet, J. Bossy, J. Y. Henry, and G. Lapertot, *Physica* 185-189 C (1991) 86.

[56]   J. W. Loram, K. A. Mirza, J. R. Cooper, and W. Y. Liang, *Phys. Rev. Lett.* 71 (1993) 1740.

[57]   C. C. Homes, T. Timusk, R. Liang, D. A. Bonn, and W. N. Hardy, *Phys. Rev. Lett.* 71 (1993) 1645.

[58]   D. N. Basov, T. Timusk, B. Dabrowski, and J. D. Jorgensen, *Phys. Rev. B* 50 (1994) 3511.

[59]  A. V. Puchkov, P. Fournier, D. N. Basov, T. Timusk, A. Kapitulnik, and N. N. Kolesnikov, *Phys. Rev. Lett.* 77 (1996) 3212.

[60]  H. Ding, T. Yokoya, J. C. Campuzano, T. Takahashi, M. Randeria, M. R. Norman, T. Mochiku, K. Kadowaki, and J. Giapintzakis, *Nature*  382  (1996) 51.

[61]  A. G. Loeser, Z.-X. Shen, D. S. Dessau, D. S. Marshall, C. H. Park, P. Fournier, and A. Kapitulnik, *Science*  273 (1996) 325.

[62]  Y. Suzumura, Y. Hasegawa, and H. Fukuyama, *J. Phys. Soc. Japan*  57 (1988) 2768; H. Fukuyama, *Prog. Theoret. Phys. Suppl.* 108 (1992) 287.

[63]  N. Nagaosa and P. A. Lee, *Phys. Rev. B* 45 (1992) 966.

[64]  M. Randeria, N. Trivedi, A. Moreo, and R. T. Scalettar, *Phys. Rev. Lett.* 69 (1992) 2001.

[65]  V. J. Emery and S. A. Kivelson, *Nature*  374 (1995) 434.

[66]  Y. J. Uemura, in *Physics of the 10th Anniversary HTS Workshop on Physics, Materials, and Applications*, B. Batlogg, C. W. Chu, W. K. Chu, D. U. Gubser, and K. A. Müller, eds. (World Scientific, Singapore, 1996), p. 68.

[67]  S.-C. Zhang, *Science* 275 (1997) 1089.

[68]  See, for example, D. L. Cox and M. B. Maple, *Physics Today*, February (1995) 32.

[69]  See, for example, B. G. Levi, *Physics Today* , May (1993) 17; January (1996) 19; and references therein.

[70]  B. Batlogg, in *High Temperature Superconductivity; Proc. Los Alamos Symp. 1989*, K. S. Bedell, D. Coffey, D. E. Meltzer, D. Pines, and J. R. Schrieffer, eds. (Addison-Wesley, Redwood City, 1990), p. 37.

[71]  W. N. Hardy, D. A. Bonn, D. C. Morgan, R. Liang, and K. Zhang, *Phys. Rev. Lett.* 70 (1993) 3999.

[72]  D. A. Bonn, R. Liang, T. M. Riseman, D. J. Baar, D. C. Morgan, K. Zhang, P. Dosanjh, T. L. Duty, A. MacFarlane, G. D. Morris, J. H. Brewer, W. N. Hardy, C. Kallin, and A. J. Berlinsky, *Phys. Rev. B* 47 (1993) 11314.

[73]  J. A. Martindale, S. E. Barrett, K. E. O'Hara, C. P. Schlichter, W. C. Lee, and D. M. Ginsberg, *Phys. Rev. B* 47 (1993) 9155.

[74]  K. A. Moler, D. J. Baar, J. S. Urbach, R. Liang, W. N. Hardy, and A. Kapitulnik, *Phys. Rev. Lett.* 73 (1994) 2744.

[75]  H. Aubin, K. Behnia, M. Ribault, R. Gagnon, and L. Taillefer, *Phys. Rev. Lett.* 78 (1997) 2624.

[76]   Z.-X. Shen, D. S. Dessau, B. O. Wells, D. M. King, W. E. Spicer, A. J. Arko, D. Marshall, L. W. Lombardi, A. Kapitulnik, P. Dickenson, S. Doniach, J. DiCarlo, T. Loeser, and C. H. Park, *Phys. Rev. Lett.* 70 (1993) 1553.

[77]   D. Mandrus, J. Hartge, C. Kendziora, L. Mihaly, and L. Forro, *Europhys. Lett.* 22 (1990) 460.

[78]   D. Coffey and L. Coffey, *Phys. Rev. Lett.* 70 (1993) 1529.

[79]   T. P. Deveraux, D. Einzel, B. Stadlober, R. Hackl, D. H. Leach, and J. J. Neumeier, *Phys. Rev. Lett.* 72 (1994) 396.

[80]   D.-H. Wu, J. Mao, S. N. Mao, J. L. Peng, X. X. Xi, R. L. Greene, and S. M. Anlage, *Phys. Rev. Lett.* 70 (1993) 85.

[81]   S. M. Anlage, D.-H. Wu, J. Mao, S. N. Mao, X. X. Xi, T. Venkatesan, J. L. Peng, and R. L. Greene, *Phys. Rev. B* 50 (1994) 523.

[82]   B. Stadlober, G. Krug, R. Nemetschek, R. Hackl, J. L. Cobb, and J. T. Markert, *Phys. Rev. Lett.* 74 (1995) 4911.

[83]   D. A. Wollman, D. J. Van Harlingen, W. C. Lee, D. M. Ginsberg, and A. J. Leggett, *Phys. Rev. Lett.* 71 (1993) 2134.

[84]   D. A. Brauner and H. R. Ott, *Phys. Rev. B* 50 (1994) 6530.

[85]   A. Mathai, Y. Gin, R. C. Black, A. Amar, and F. C. Wellstood, *Phys. Rev. Lett.* 74 (1995) 4523.

[86]   D. A. Wollman, D. J. Van Harlingen, J. Giapintzakis, and D. M. Ginsberg, *Phys. Rev. Lett.* 74 (1995) 797.

[87]   J. H. Miller, Q. Y. Ying, Z. G. Zou, N. Q. Fan, J. H. Xu, M. F. Davis, and J. C. Wolfe, *Phys. Rev. Lett.* 74 (1995) 2347.

[88]   C. C. Tsuei, J. R. Kirtley, C. C. Chi, L.-S. Yu-Jahnes, A. Gupta, T. Shaw, J. Z. Sun, and M. B. Ketchen, *Phys. Rev. Lett.* 72 (1994) 593.

[89]   J. R. Kirtley, C. C. Tsuei, J. Z. Sun, C. C. Chi, L. S. Yu-Jahnes, A. Gupta, M. Rupp, and M. B. Ketchen, *Nature* 373 (1995) 225.

[90]   C. C. Tsuei, J. R. Kirtley, M. Rupp, J. Z. Sun, A. Gupta, M. B. Ketchen, C. A. Wang, Z. F. Ren, J. H. Wang, and M. Bhushan, *Science* 271 (1996) 329.

[91]   A. G. Sun, D. A. Gajewski, M. B. Maple, and R. C. Dynes, *Phys. Rev. Lett.* 72 (1994) 2267.

[92]   A. G. Sun, S. H. Han, A. S. Katz, D. A. Gajewski, M. B. Maple, and R. C. Dynes, *Phys. Rev. B* 52 (1995) R15731.

[93]   J. Lesueur, M. Aprili, A. Goulan, T. J. Horton, and L. Dumoulin, *Phys. Rev. B* 55 (1997) 3308.

[94]   A. G. Sun, A. Truscott, A. S. Katz, R. C. Dynes, B. W. Veal, and C. Gu, *Phys. Rev. B* 54 (1996) 6734.

[95]   R. Kleiner, A. S. Katz, A. G. Sun, R. Summer, D. A. Gajewski, S. H. Han, S. I. Woods, E. Dantsker, B. Chen, K. Char, M. B. Maple, R. C. Dynes, and J. Clarke, *Phys. Rev. Lett.* 76 (1996) 2161.

[96]   P. Chaudhari and S.-Y. Lin, *Phys. Rev. Lett.* 72 (1994) 1084.

[97]   K. A. Kouznetsov, A. G. Sun, B. Chen, A. S. Katz, S. R. Bahcall, J. Clarke, R. C. Dynes, D. A. Gajewski, S. H. Han, M. B. Maple, J. Giapintzakis, J.-T. Kim, and D. M. Ginsberg, to appear in *Phys. Rev. Lett.*, 1997.

[98]   H. Srikanth, B. A. Willemsen, T. Jacobs, S. Sridhar, A. Erb, E. Walker, and R. Flükiger, *Phys. Rev. B* 55 (1997) R14733.

[99]   A. Erb, E. Walker, and R. Flükiger, *Physica C* 245 (1995) 245.

[100]  J. M. Rowell, *Solid State Commun.* 102 (1997) 269.

[101]  R. Nagarajan, C. Mazumdar, Z. Hossain, S. K. Dhar, K. V. Gopalakrishnan, L. C. Gupta, C. Godart, B. D. Padalia, and R. Vijayaraghavan, *Phys. Rev. Lett.* 72 (1994) 274.

[102]  R. J. Cava, H. Takagi, B. Batlogg, H. W. Zandbergen, J. J. Krajewski, W. F. Peck, Jr., R. B. van Dover, R. J. Felder, K. Mizuhashi, J. O. Lee, H. Eisaki, and S. Uchida, *Nature* 367 (1994) 252.

[103]  For a review, see *Superconductivity in Ternary Compounds II; Superconductivity and Magnetism*, M. B. Maple and Ø. Fischer, eds. (Springer, Berlin, Heidelberg, New York, 1982).

[104]  B. K. Cho, P. C. Canfield, and D. C. Johnston, *Phys. Rev. B* 52 (1995) R3844.

[105]  R. J. Cava, H. Takagi, B. Batlogg, H. W. Zandbergen, J. J. Krajewski, W. F. Peck, Jr., R. B. van Dover, R. J. Felder, T. Siegrist, K. Mizuhashi, J. O. Lee, H. Eisaki, S. A. Carter, and S. Uchida, *Nature* 367 (1994) 146.

[106]  J. L. Sarrao, M. C. de Andrade, J. Herrmann, S. H. Han, Z. Fisk, M. B. Maple, and R. J. Cava, *Physica C* 229 (1994) 65.

[107]  Y. Maeno, H. Hashimoto, K. Yoshida, S. Nishizaki, T. Fujita, J. G. Bednorz, and P. Lichtenberg, *Nature* 372 (1994) 532.

[108]  T. M. Rice and M. Sigrist, *J. Phys., Condens. Matter* 7 (1995) 643.

[109]  A. F. Hebard, *Physics Today*, November (1992) 26; and references therein.

[110]  S. Kondo, D. C. Johnston, C. A. Swenson, F. Borsa, A. V. Mahajan, L. L. Miller, T. Gu, A. I. Goldman, M. B. Maple, D. A. Gajewski, E. J. Freeman, N. R. Dilley, R. P. Dickey, J. Merrin, K. Kojima, G. M. Luke, Y. J. Uemura, O. Chmaissem, and J. D. Jorgensen, *Phys. Rev. Lett.* 78 (1997) 3729.

[111]   M. B. Maple, M. C. de Andrade, J. Herrmann, R. P. Dickey, N. R. Dilley, and S. Han, *J. Alloys and Compounds*  250 (1997) 585.

[112]   D. C. Johnston, *J. Low Temp. Phys.* 25 (1976) 145.

[113]   M. Takano, *Physica C*  263 (1996) 468.

[114]   S. Maekawa, *Science*  273 (1996) 1515.

[115]   M. Uehara, T. Nagata, J. Akimitsu, H. Takahashi, N. Môri, and K. Kinoshita, *J. Phys. Soc. Japan* 65 (1996) 2764.

**Table caption**

Table 1.

(a) Some important classes of cuprate superconductors and the maximum value of $T_c$ observed in each class.

(b) Examples of the abbreviated designations (nicknames) used to denote cuprate materials.

**Figure captions**

Fig. 1.

Maximum superconducting critical temperature $T_c$ vs date.

Fig. 2.

Temperature-dopant concentration (T-x) phase diagram delineating the regions of superconductivity and antiferromagnetic ordering of the $Cu^{2+}$ ions for the hole-doped $La_{2-x}Sr_xCuO_4$ and electron-doped $Nd_{2-x}Ce_xCuO_{4-y}$ systems. AFM = antiferromagnetic phase, SG = spin-glass phase, and SC = superconducting phase. After Reference [19].

Fig. 3.

Crystal structures of $La_2CuO_4$ (T-phase) and $Ln_2CuO_4$ (Ln = Pr, Nd, Sm, Eu, Gd; T'-phase) parent compounds. From Reference [19].

Fig. 4.

(a) In-plane electrical resistivity $\rho_{ab}$ vs temperature T curves for $Y_{1-x}Pr_xBa_2Cu_3O_{7-\delta}$ ($0 \le x \le 0.51$) single crystals. From Reference [36].

(b) Temperature T vs Pr concentration x phase diagram for tthe $Y_{1-x}Pr_xBa_2Cu_3O_{7-\delta}$ system, delineating metallic, superconducting, insulating, and antiferromagnetically ordered regions. From References [36,37].

Fig. 5.

(a)  Temperature dependence of the Hall carrier number $n_H = V/eR_H$ vs temperature T for $Y_{1-x}Pr_xBa_2Cu_3O_{7-\delta}$ single crystals with different x values.  From Reference [36].

(b)  Cotangent of the Hall angle $\cot(\theta_H)$ vs $T^2$ for $Y_{1-x}Pr_xBa_2Cu_3O_{7-\delta}$ single crystals with different x values.  Inset: Slope A and intercept B vs T.  From Reference [36].

Fig. 6.

(a)  In-plane resistivity $\rho_{ab}(T)$ and (b) out-of-plane resisitivity $\rho_c(T)$ for $Y_{1-x}Pr_xBa_2Cu_3O_{7-\delta}$ single crystals.  The solid lines in (b) are fits to the data using a model described in Reference [43].  Inset to Fig. 6(a): The configuration of leads used in the measurements.  Inset to Fig. 6(b): Anisotropy $\rho_c/\rho_{ab}$ vs Pr concentration x at different temperatures.  The solid lines are guides to the eye.  From Reference [43].

Fig. 7.

Momentum and temperature dependence of the energy gap estimated from leading edge shifts of ARPES spectra for BSCCO-2212.  (a) k-dependence of the gap in the $T_c = 87$ K, 83 K, and 10 K samples, measured at 14 K.  The inset shows the Brillouin zone with a large Fermi surface (FS) closing the $(\pi,\pi)$ point, with the occupied region shaded.  (b) Temperature dependence of the maximum gap in a near-optimal $T_c = 87$ K sample (circles), and underdoped $T_c = 83$ K (squares) and $T_c = 10$ K (triangles) samples.  From Reference [60].

Fig. 8.

Schematic phase diagram of BSCCO-2212 as a function of doping. The filled symbols represent the $T_c$'s determined from magnetic susceptibility measurements. The open symbols are the T*'s at which the pseudogap closes derived from the data shown in Fig. 7. From Reference [60].

Fig. 9.

Generic temperature T - dopant concentration x phase diagram for cuprates (schematic). The solid lines labelled $T_N$ and $T_c$ delineate the antiferromagnetic (AFM) and superconducting regions, respectively. The "hatched" line denoted T* represents the crossover into the pseudogap state.

Fig. 10.

Schematic temperature T - dopant concentration x phase diagram for the cuprates. The dashed line labelled T* represents the temperature below which some type of local pairing occurs leading to a suppression of low energy excitations and the formation of the pseudogap. The solid line labelled $T_\phi$ denotes the temperature below which phase coherence develops, resulting in superconductivity. The dark solid line labelled $T_c$ delineates the superconducting region.

Fig. 11.

Fermi surface gap functions and densities of states of a superconductor with tetragonal symmetry for various pairing symmetries. The gap functions in the $k_z = 0$ plane (top) are represented by the light solid lines; distance from the Fermi surface (dark solid lines) gives the amplitude, a positive value being outside the Fermi surface, a negative value inside. The corresponding density of states for one-quasiparticle excitations N(E) is shown below each gap function, with $N_o$ the normal state value. Gap node surfaces are represented by the dashed lines. Left: The classic s-wave case, where the gap function is constant, with value $\Delta$. This gives rise to a square-root singularity in N(E) at energy $E = \Delta$. Middle: The extended s-wave case derives from pairs situated on nearest-neighbor square lattice sites in real space, with an approximate k-space form of $\cos(k_x a) + \cos(k_y a)$. For the Fermi surface shown here, the gap function has lines of nodes running out of the page. Right: A d-wave function of $x^2 - y^2$ symmetry. The extended s-wave and d-wave functions shown here each have a linear density of states up to order $\Delta$, which measures the maximum gap amplitude about the Fermi surface. From Reference [68].

Fig. 12.

Josephson critical current $I_c$ vs magnetic field B for c-axis Josephson tunneling between an untwinned $YBa_2Cu_3O_{7-\delta}$ single crystal and a Pb counterelectrode for B‖a (upper curve) and B‖b (lower curve). The upper curve is offset by 0.5 mA along the y-axis. From Reference [92].

Fig. 13.

Magnetic field dependence of the critical current density for a range of short sample YBCO conductors produced using either IBAD or RABIT substrates. These data are compared with typical values obtained for NbTi and $Nb_3Sn$ wires at 4.2 K and for BSCCO, oxide-powder-in-tube wires at 77K. From Reference [14].

**Table 1.** (a) Some important classes of cuprate superconductors and the maximum value of $T_c$ observed in each class. (b) Examples of the abbreviated names (nicknames) used to denote cuprate materials.

**(a)**

| Material | | Max. $T_c$(K) |
|---|---|---|
| • $La_{2-x}M_xCuO_4$; M = Ba, Sr, Ca, Na | | ~40 |
| • $Ln_{2-x}M_xCuO_{4-y}$; Ln = Pr, Nd, Sm, Eu; M = Ce, Th | | ~25 |
| • $YBa_2Cu_3O_{7-\delta}$ | | 92 |
| • $LnBa_2Cu_3O_{7-\delta}$<br>Ce, Tb do not form phase<br>Pr forms phase; neither metallic nor SC'ing | | ~95 |
| • $RBa_2Cu_4O_8$ | | ~80 |
| • $Bi_2Sr_2Ca_{n-1}Cu_nO_{2n+4}$ | (n = 1,2,3,4) | (n = 3) 110 |
| • $TlBa_2Ca_{n-1}Cu_nO_{2n+3}$ | (n = 1,2,3,4) | (n = 4) 122 |
| • $Tl_2Ba_2Ca_{n-1}Cu_nO_{2n+4}$ | (n = 1,2,3,4) | (n = 3) 122 |
| • $HgBa_2Ca_{n-1}Cu_nO_{2n+2}$ | (n = 1,2,3,4) | (n = 3) 133 |

**(b)**

| Material | Nickname | $T_c$(K) |
|---|---|---|
| • $YBa_2Cu_3O_{7-\delta}$ | YBCO; YBCO–123; Y–123 | 92 |
| • $Bi_2Sr_2Ca_2Cu_3O_{10}$ | BSCCO; BSCCO–2223; Bi–2223 | 110 |
| • $Tl_2Ba_2Ca_2Cu_3O_{10}$ | TBCCO; TBCCO–2223; Tl–2223 | 122 |
| • $HgBa_2Ca_2Cu_3O_8$ | HBCCO; HBCCO–1223; Hg–1223 | 133 |
| • $La_{1.85}Sr_{0.15}CuO_4$ | LSCO | 40 |
| • $Nd_{1.85}Ce_{0.15}CuO_{4-y}$ | NCCO | 25 |

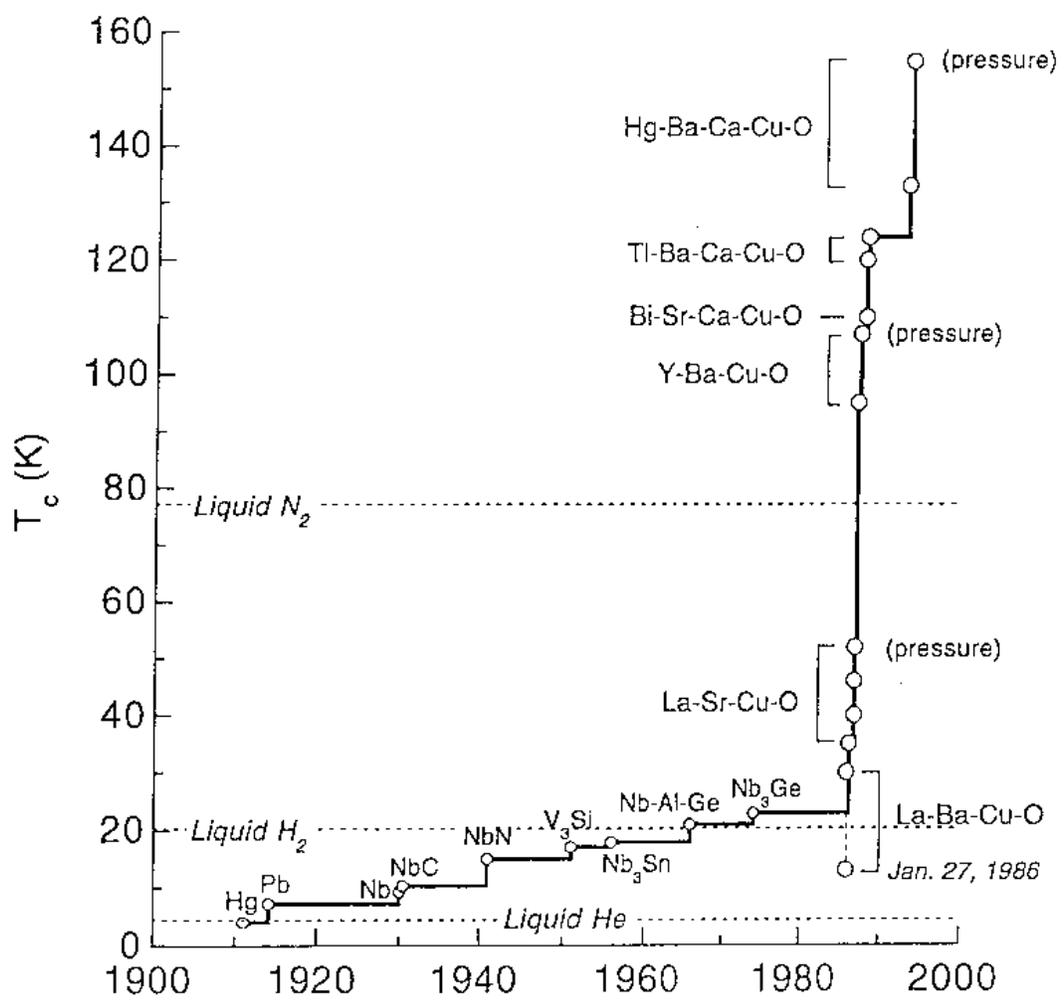

Fig. 1

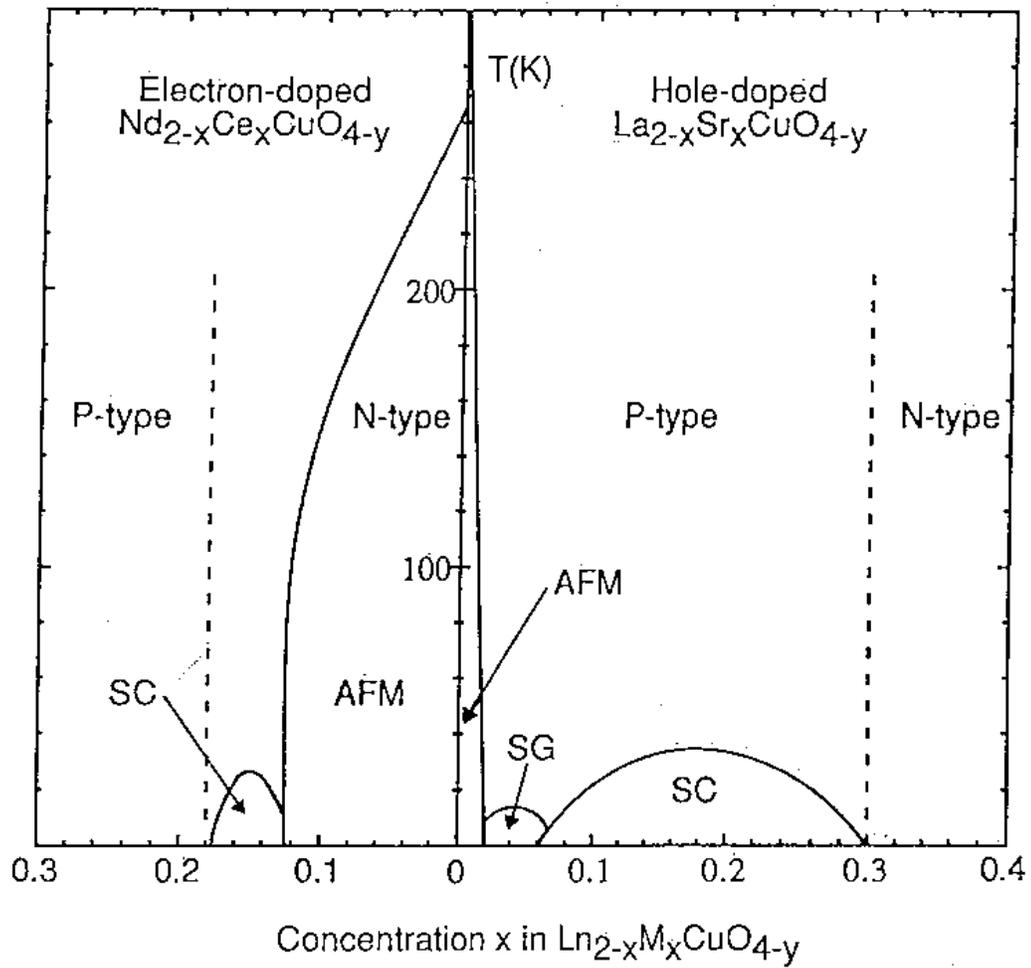

ELECTRON-HOLE SYMMETRY (QUALITATIVE)

Metallic ←——— Insulating ———→ Metallic

Fig. 2

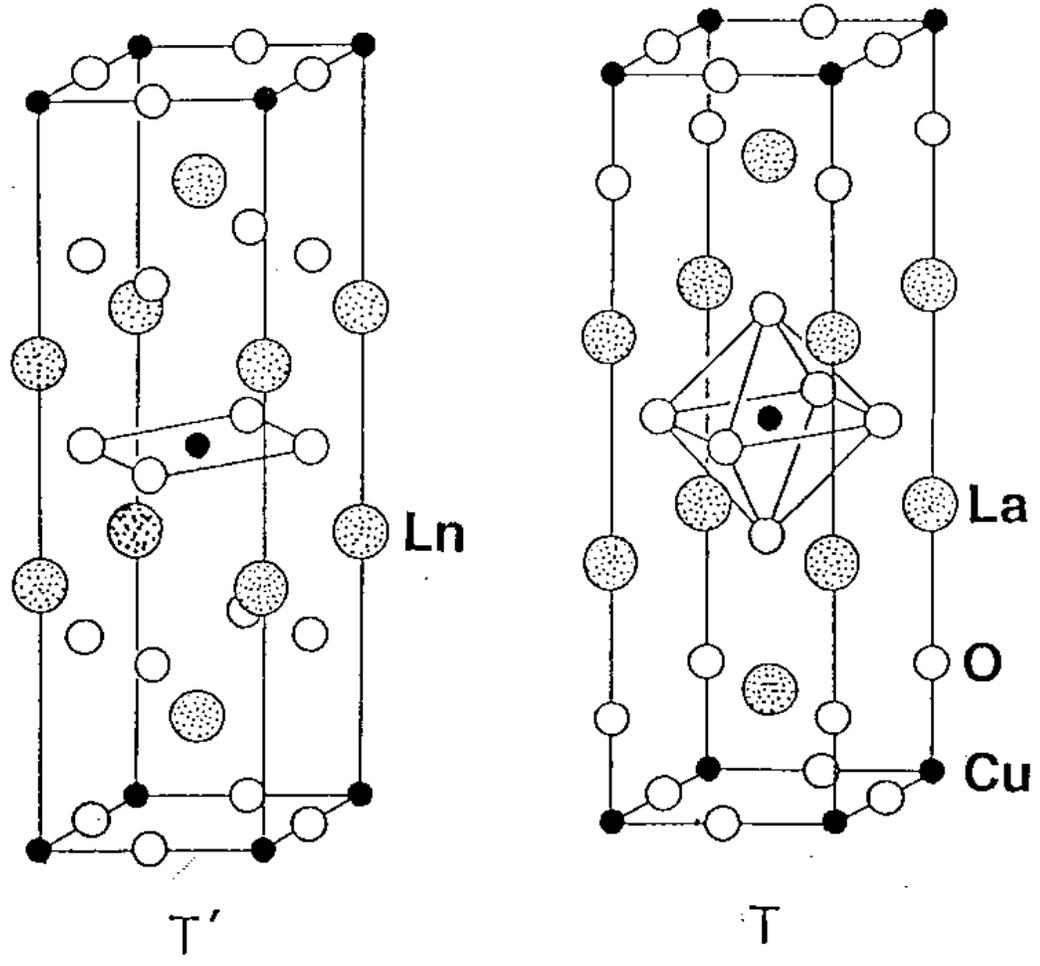

T'                    T

Ln                   La

                     O

                     Cu

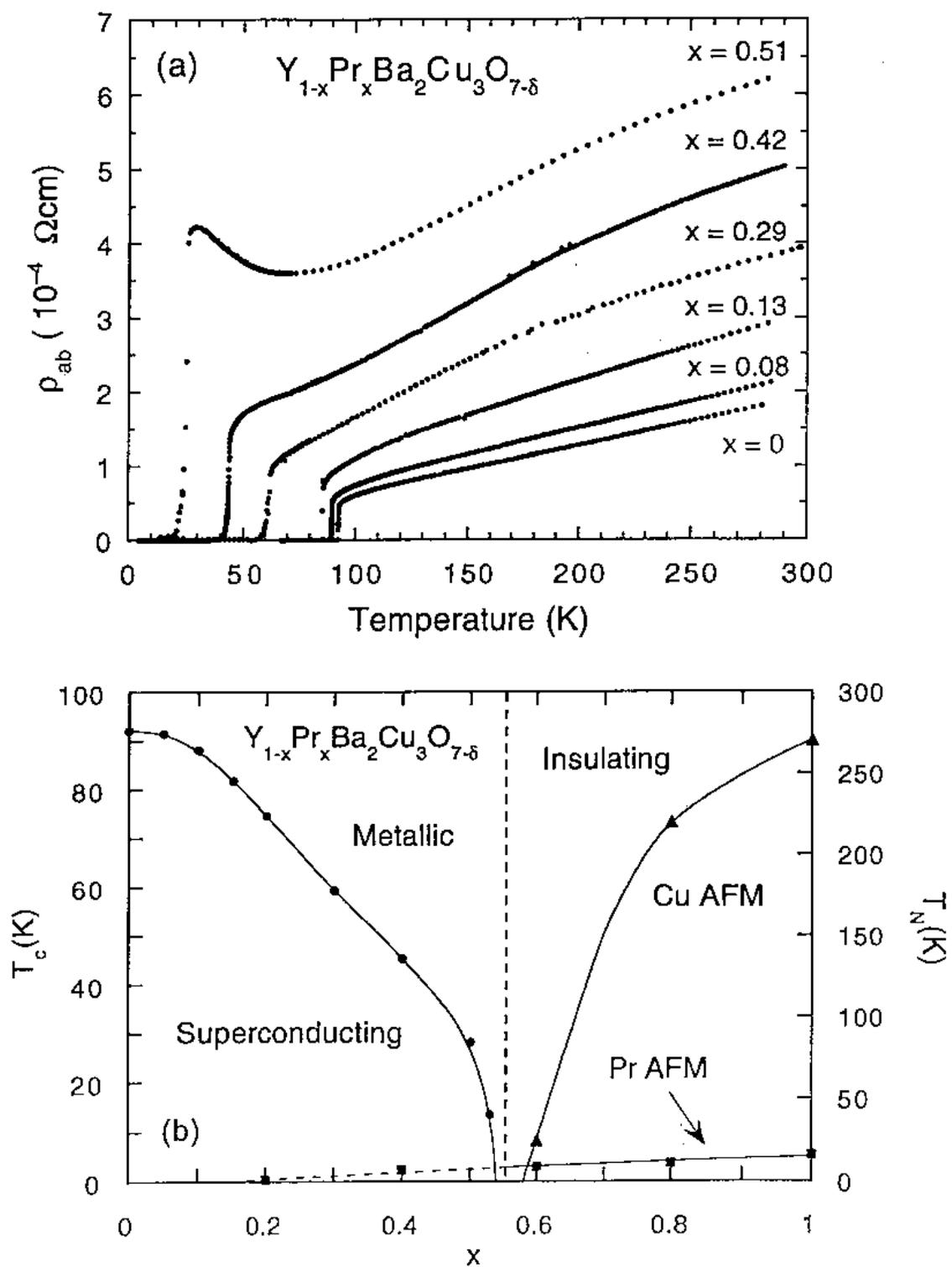

Fig. 4

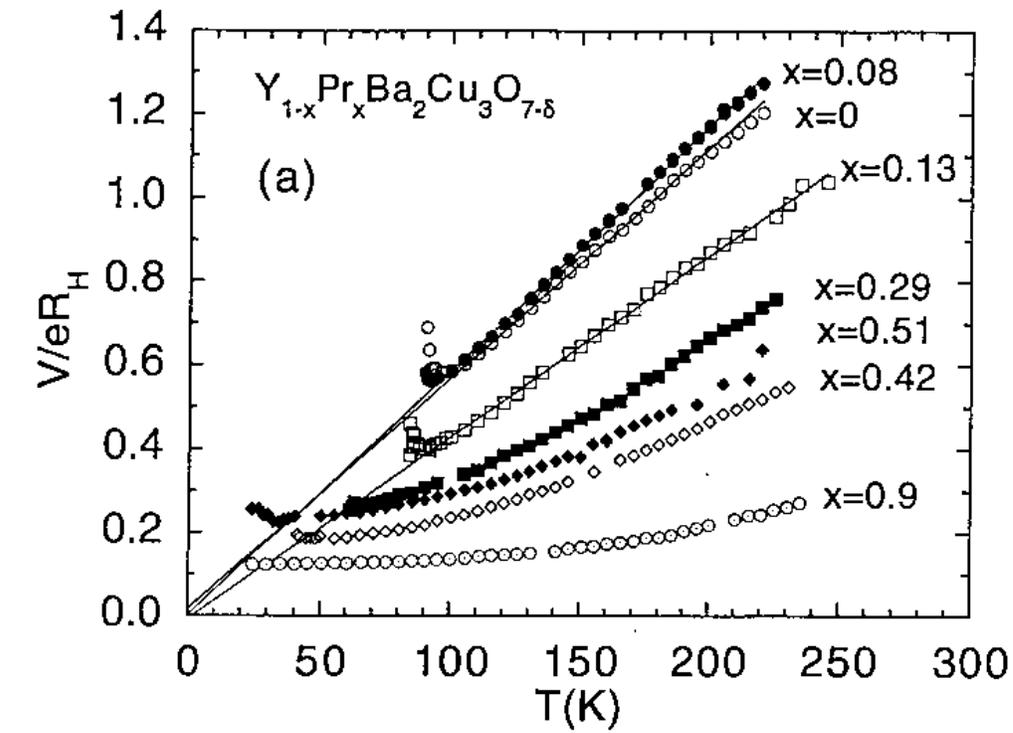

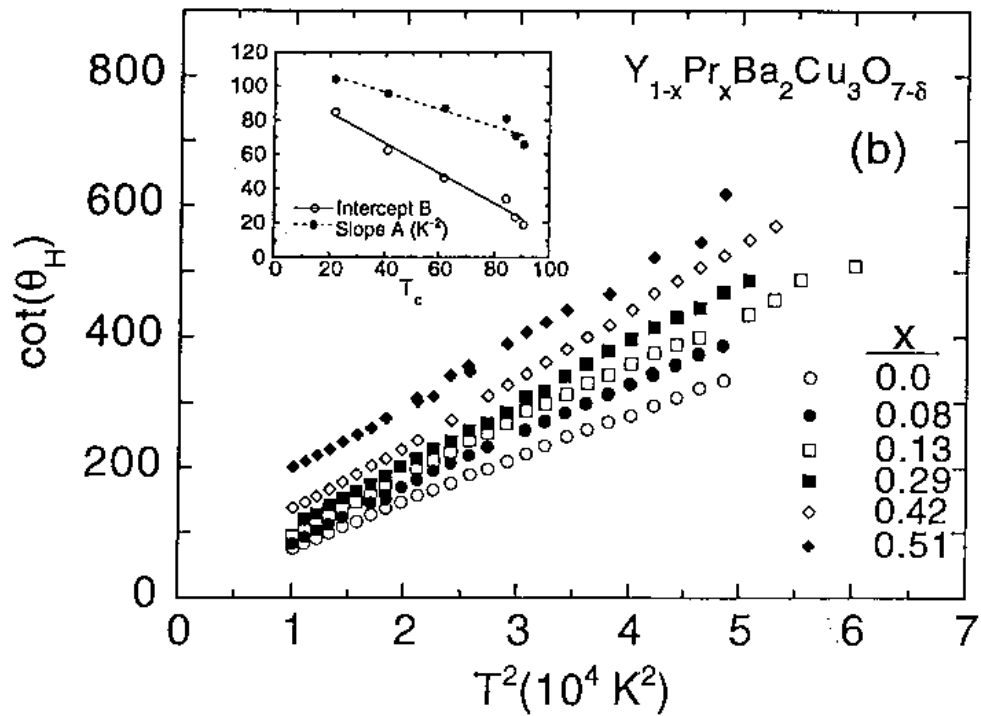

Fig. 5

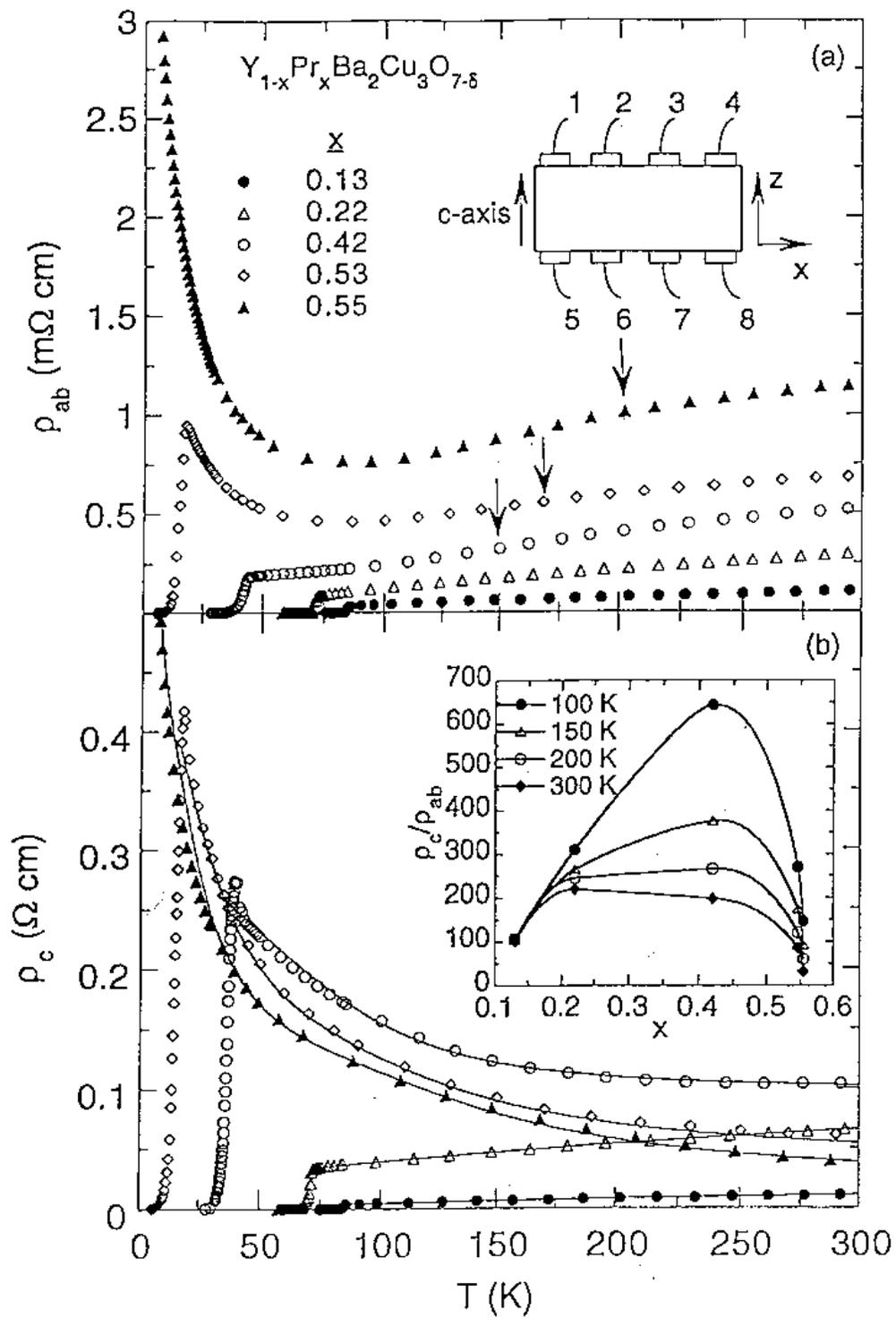

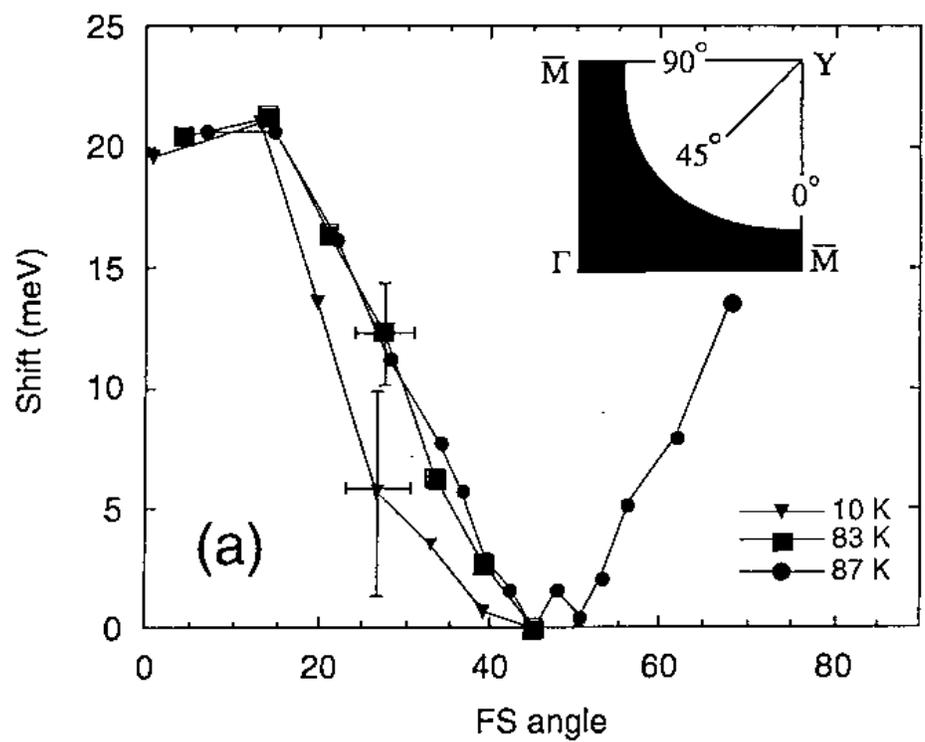

(a)

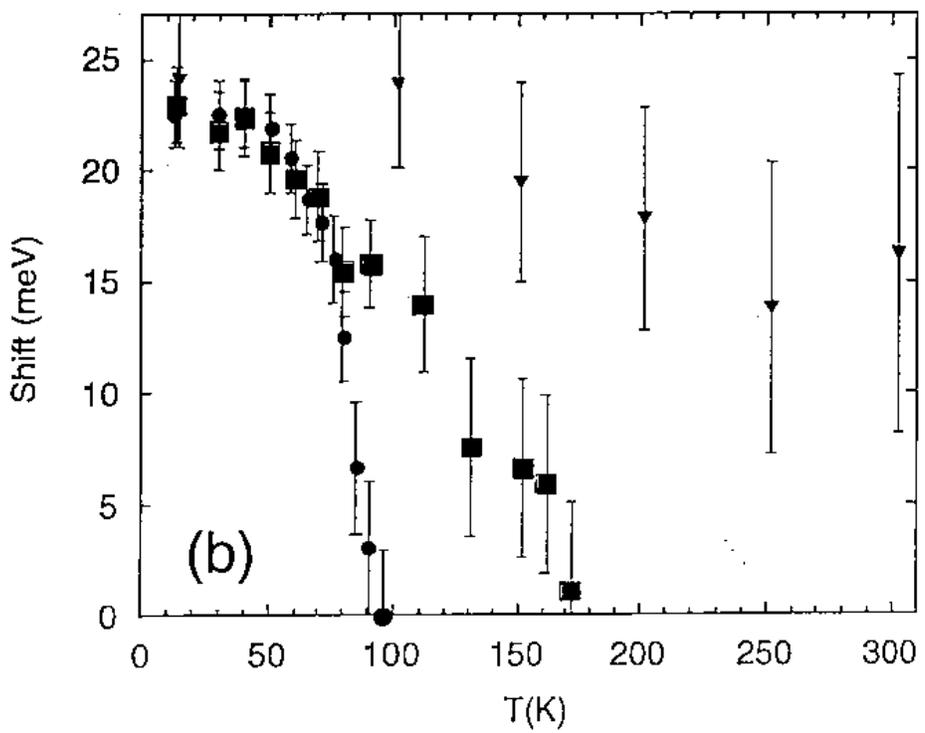

(b)

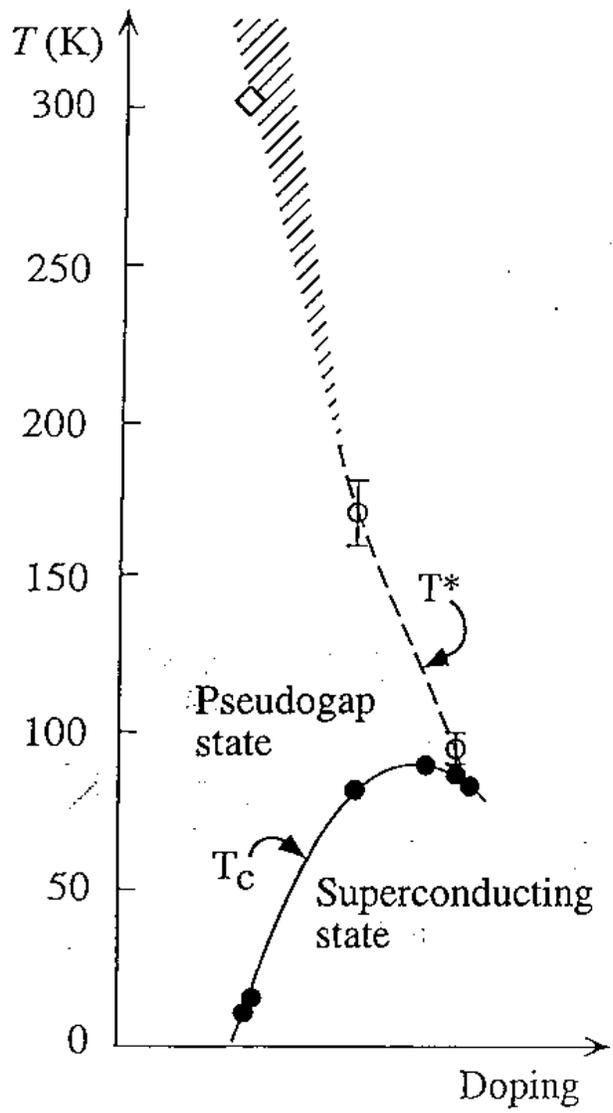

Fig. 8

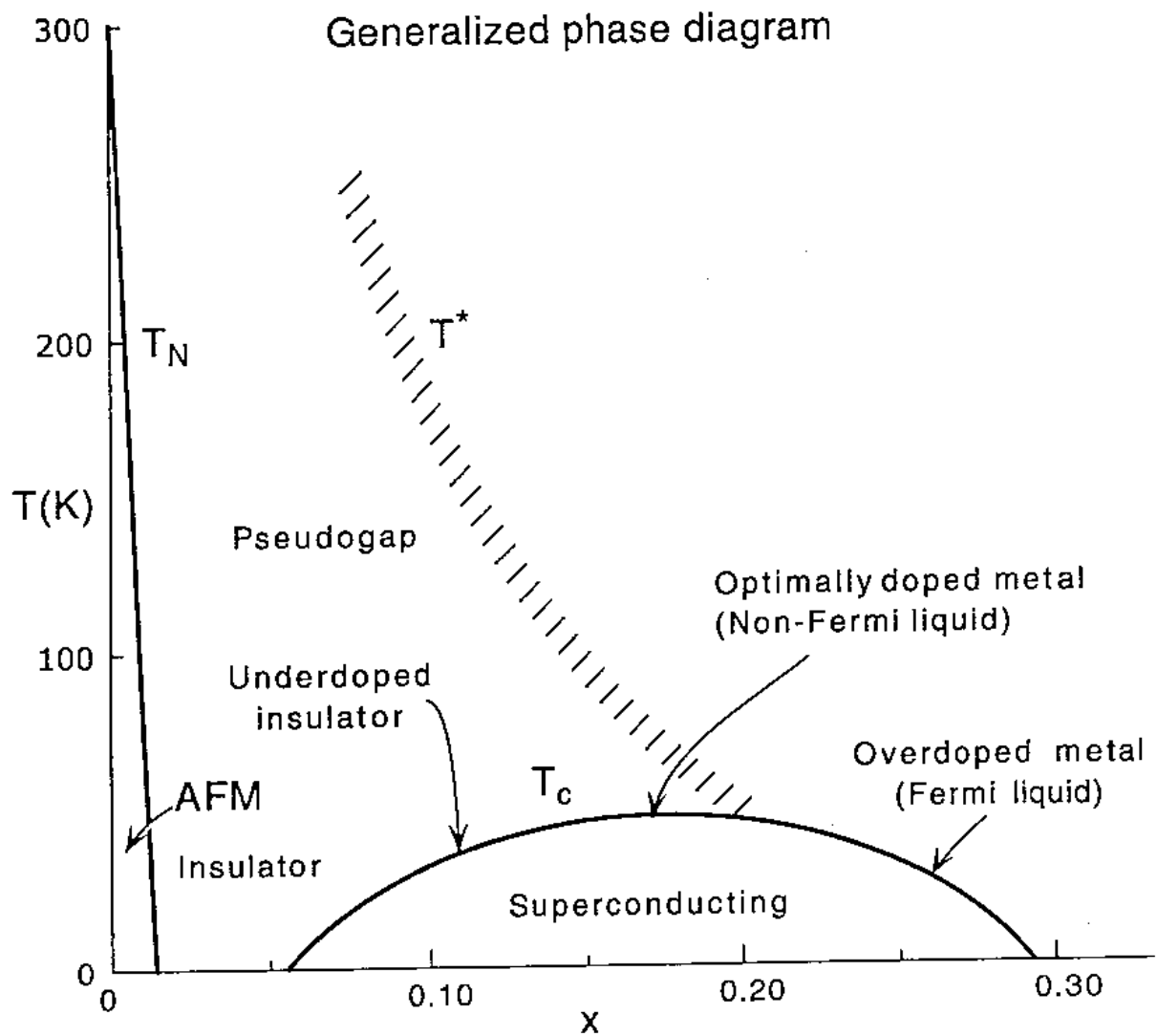

Generalized phase diagram

Fig. 9

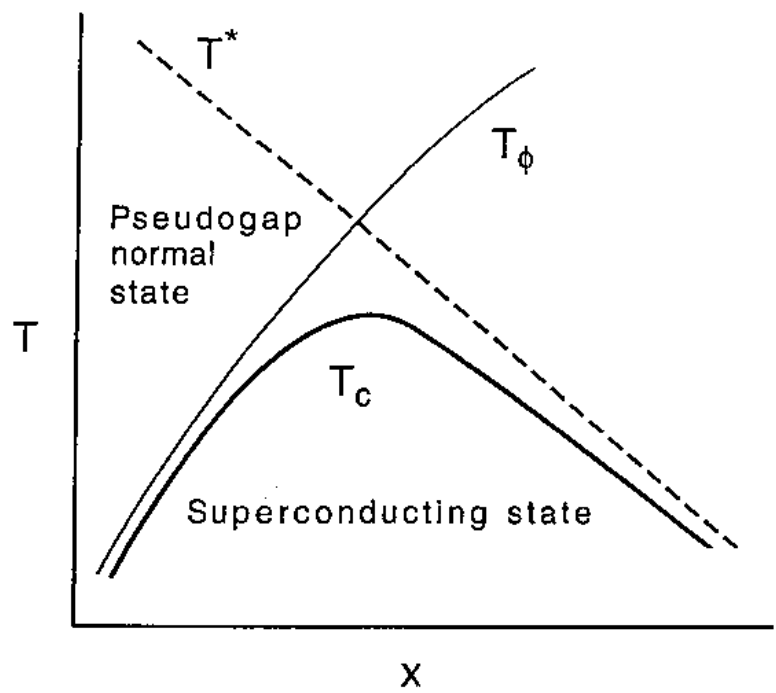

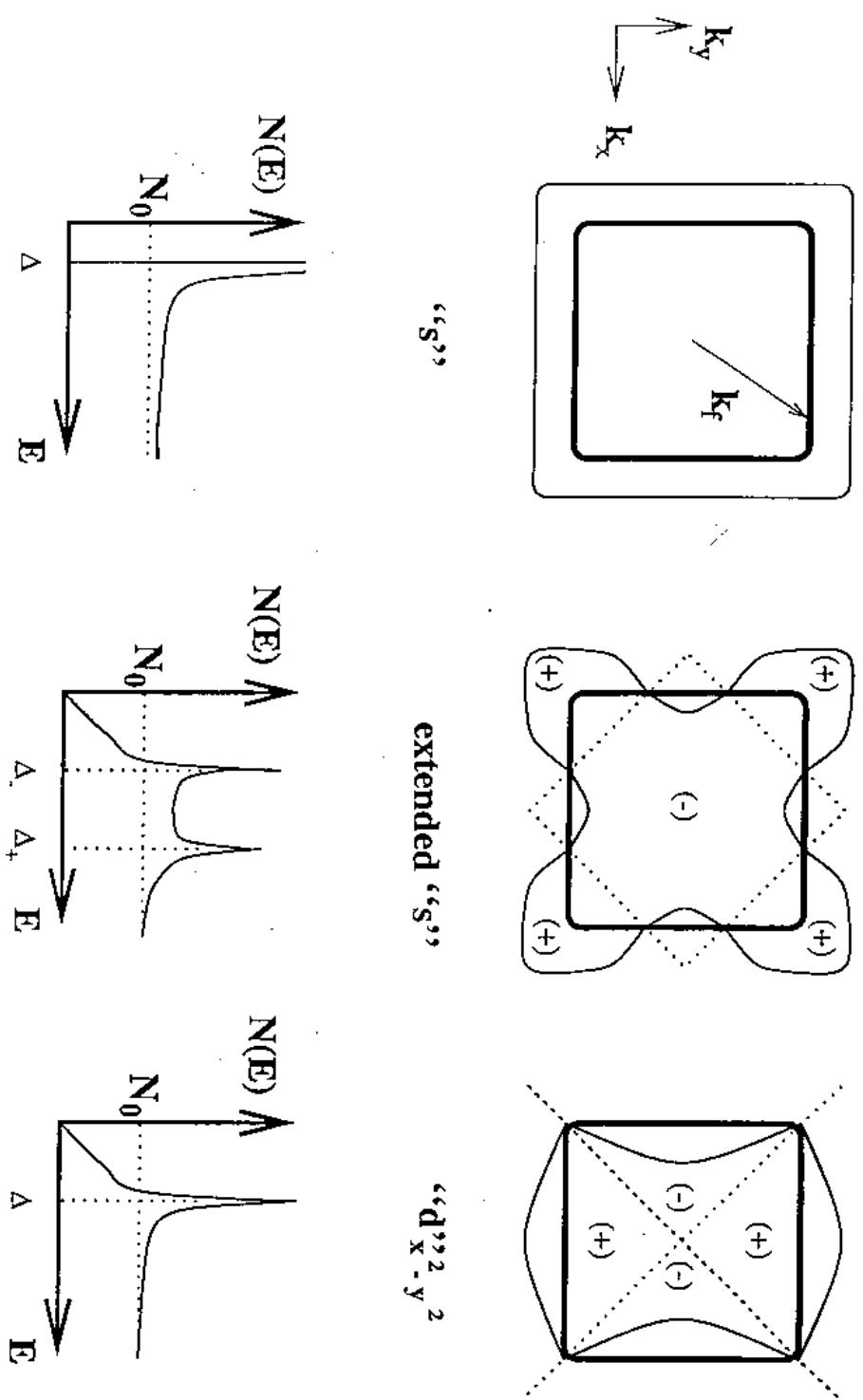

Fig. 11

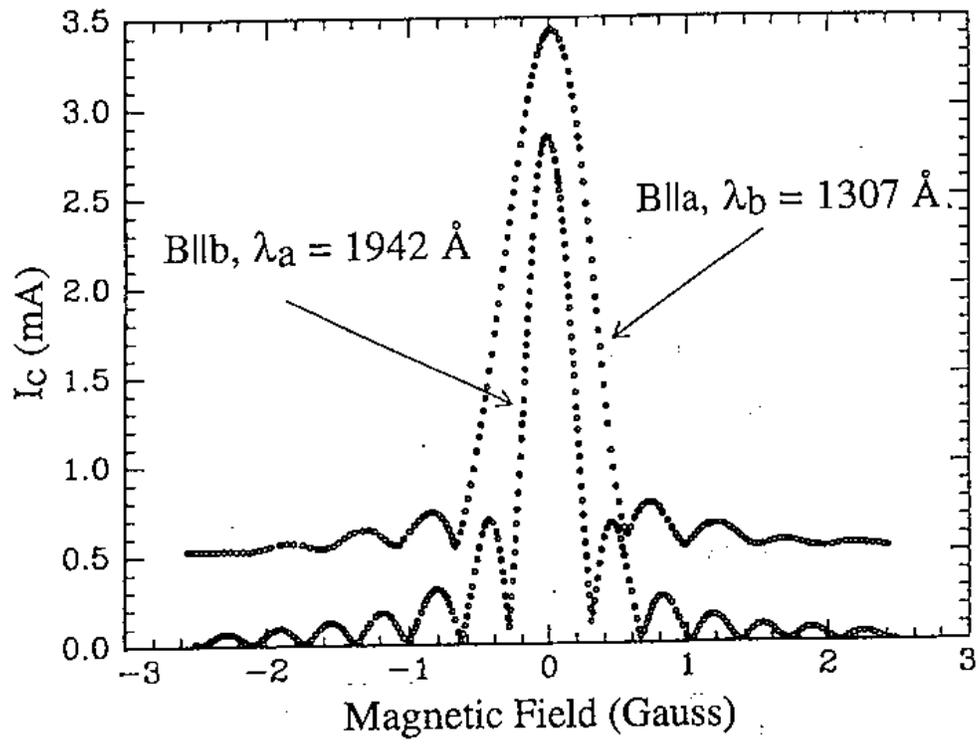

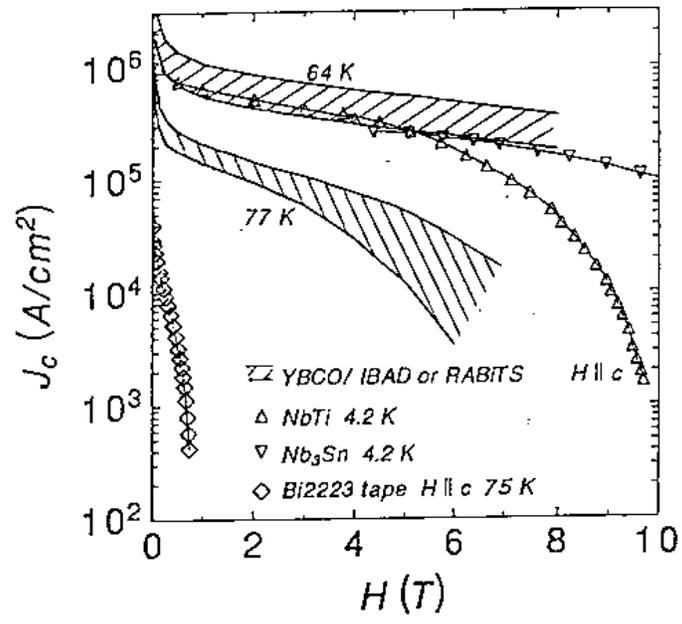

Fig. 13